\documentclass[conference,10pt]{IEEEtran}

\usepackage{subfig}

\usepackage{epsfig,amsmath,amssymb,epsf,cite,algorithm,algorithmic}
\usepackage{graphicx}

\newcommand{\argmax}{\mathop{\mathrm{argmax}}}

\usepackage{stmaryrd}
\usepackage{amssymb}
\usepackage{tipa}
\usepackage{setspace}

\graphicspath{{./figs/}}

\def\b0{{\pmb{0}}}

 \def\bh{{\textbf{h}}}  
   
   \def\bw{{\textbf{w}}}
   \def\bx{{\textbf{x}}}

\def\bA{{\textbf{A}}} \def\bH{{\textbf{H}}}

   \def\bX{{\textbf{X}}}
   \def\bY{{\textbf{Y}}}
   \def\bZ{{\textbf{Z}}}
 \def\bN{{\textbf{N}}}

\IEEEoverridecommandlockouts

\begin{document}
\title{Sparse Channel Estimation for Massive MIMO with 1-bit Feedback per Dimension}
\author{Zhiyi Zhou, Xu Chen, Dongning Guo and Michael L. Honig \\
Department of Electrical Engineering and Computer Science\\
 Northwestern University, Evanston, IL, 60208, USA

\thanks{This work was partially supported by a gift from Futurewei Technologies.}
}
%


\maketitle

\begin{abstract}
In massive multiple-input multiple-output (MIMO) systems, acquisition of the channel state information at the transmitter side (CSIT) is crucial. In this paper, a practical CSIT estimation scheme is proposed for frequency division duplexing (FDD) massive MIMO systems. Specifically, each received pilot symbol is first quantized to one bit per dimension at the receiver side and then the quantized bits are fed back to the transmitter. A joint one-bit compressed sensing algorithm is implemented at the transmitter to recover the channel matrices. The algorithm leverages the hidden joint sparsity structure in the user channel matrices to minimize the training and feedback overhead, which is considered to be a major challenge for FDD systems. Moreover, the one-bit compressed sensing algorithm accurately recovers the channel directions for beamforming. The one-bit feedback mechanism can be implemented in practical systems using the uplink control channel. Simulation results show that the proposed scheme nearly achieves the maximum output signal-to-noise-ratio for beamforming based on the estimated CSIT.  
\end{abstract}

\section{Introduction}
\label{sec:intro}

Massive multiple-input multiple-output (MIMO) will be an enabling technique for next generation wireless communications due to its large degrees of freedom\cite{larsson2014massive,hoydis2013massive}. As the number of antennas at the base terminal station (BTS) increases, massive MIMO can mitigate the inter-user interference in both downlink and uplink multiuser systems with simple precoders and receivers~\cite{marzetta2010noncooperative}. This relies on the fact that the random channel vectors of the users become nearly orthogonal, and thus simply aligning the beam to the desired channel can offer a significant performance gain. To maximize the benefit of massive MIMO, perfect estimation of the channel state information at the transmitter (CSIT) is needed.

There have been a lot of work on CSIT estimation for massive MIMO systems. Some of them consider time division duplexing (TDD) systems, because channel reciprocity can be utilized for efficient channel estimation using uplink pilots. On the other hand, frequency division duplexing (FDD) dominates current wireless cellular systems, which leads to a lot of studies on obtaining CSIT for massive MIMO with FDD recently \cite{marzetta2013special,choi2014downlink,kuo2012compressive,gao2015spatially,rao2014distributed,zhu2016compressive}. In this paper, we focus on a downlink training framework for FDD massive MIMO systems.  

In FDD systems, the interference caused by pilot contamination diminishes with increasing number of antennas at the BTS. The major drawback of FDD systems is that CSIT estimation requires a large amount of training overhead and feedback overhead, which usually scale linearly in the number of transmitting antennas. In order to reduce the training overhead, the temporal and spatial correlations of the channels have been exploited  in~\cite{choi2014downlink}. In addition, it is shown from many experimental studies of massive MIMO channels that the channel matrices are sparse in the angular domain due to limited scattering around the BTS \cite{Zhou2008Experimental,Kyritsi2003Corelation,Kaltenberger2008Corelation,Hoydis2012Channel,Gao2011Linear}. Therefore, such hidden sparsity structure has been utilized in \cite{kuo2012compressive,gao2015spatially,rao2014distributed,zhu2016compressive} by applying compressed sensing techniques to estimate CSIT with significantly reduced training overhead. 

However, most of the sparsity-inspired approaches assume that receivers perfectly feed back the analog received signal to the BTS~\cite{gao2015spatially,rao2014distributed,zhu2016compressive}. One possible implementation is to quantize the received signal and feed back the quantized symbols to the BTS. It is known that direct quantization of the received symbols involves high complexity and inevitably requires many bits per symbol to achieve good performance~\cite{love2004value}. Although the sparsity-inspired approaches often require a smaller amount of training overhead, the feedback overhead may still be large.

In this paper, we propose a CSIT estimation scheme for massive MIMO systems based on distributed one-bit compressed sensing. The proposed scheme significantly reduces the feedback overhead. In particular, the received symbols are first quantized to one bit per dimension and then sent to the BTS. The BTS jointly recovers the CSIT by exploiting the hidden sparsity of the channels. The proposed scheme has several advantages compared with existing sparsity-inspired approaches. First, one-bit per dimension feedback greatly reduces the amount of feedback compared with multi-bit quantized analog feedback. Second, the channel directions can be correctly estimated, with only the sign information of the received signal. Simulation results show that beamforming based on the estimated CSIT has less than 0.5 dB signal-to-noise-ratio (SNR) degradation compared to the maximum achievable output SNR. Third, to the best of our knowledge, this paper is the first to consider the recovery of jointly sparse signals based on one-bit quantized measurements.

Throughput the paper, upper-case and lower-case boldface symbols are used to denote matrices and column vectors, respectively. In addition, $\bX^H$ denotes the Hermitian transpose of $\bX$ and ${||\bX||}_F$ denotes the Frobenius norm of $\bX$.

\section{System Model}
\label{sec:system}

Consider a multi-user massive MIMO system operating in FDD mode. The network consists of one BTS and $K$ users. Denote the set of users as $\mathcal{K} = \{1,\cdots, K\}$. The BTS is equipped with $M$ antennas and each of the users is equipped with $N$ antennas. The BTS transmits $T$ pilots in the downlink where the $t$-th ($t = 1,2,\cdots, T$) symbol is denoted as $\bx_t \in \mathbb{C}^{M\times 1}$, which satisfies the power constraint 
\begin{align}
\frac{1}{T}\sum\limits_{t=1}^{T}\bx_t^H\bx_t \leq P.
\end{align}
At the $i$-th user, the received pilots during $T$ channel uses can be expressed as 
\begin{equation}
\label{eq1}
\bY_{i} = \bH_i\bX+\bN_{i},
\end{equation}
where $\bH_i \in \mathbb{C}^{N\times M}$ is the quasi-static channel matrix from the BTS to user $i$, $\bX = [\bx_1,\bx_2,\cdots,\bx_T]\in \mathbb{C}^{M\times T}$ is the concatenated transmitted pilots and $\bN_{i} \in \mathbb{C}^{N\times T}$ consists of independent identically distributed complex-valued Gaussian random variables, each with distribution $\mathcal{CN}(0,1)$.

A uniform linear array (ULA) model is assumed for the antennas at the BTS and the users. Hence, the channel matrix $\bH_i$ can be represented in the angular domain as \cite{Tse2005fundamental}:
\begin{equation}
\label{eq2}
\bH_{i} = \bA_R\bH^a_{i}\bA_T^H,
\end{equation}
where $\bA_R \in \mathbb{C}^{N\times N}$ and $\bA_T \in \mathbb{C}^{M\times M}$ are the unitary matrices for the angular domain transformation at the users and BTS, respectively, and $\bH^a_{i} \in \mathbb{C}^{N\times M}$ is the channel matrix expressed in angular coordinates. The nonzero $(p,q)$-th entry of $\bH^a_{i}$ indicates that there is a path from the $q$-th angle of departure (AoD) at the BTS to the $p$-th angle of arrival (AoA) at user $i$. As shown in experimental results \cite{Zhou2008Experimental}, the angular domain matrix $\bH^{a}_{i}$ can be well modeled by a sparse matrix because of limited local scattering at the BTS side. Let $\bh_{ij}$ denote the $j$-th row of $\bH_i^a$, define $\mathcal{S}_{ij} = \{l:\bh_{ij}(l) \neq 0\}$ as the support for $\bh_{ij}$. Based on empirical measurements of the channel matrices of massive MIMO systems \cite{Kyritsi2003Corelation,Kaltenberger2008Corelation,Hoydis2012Channel,Gao2011Linear}, the channel matrices $\{\bH^a_i: i\in\mathcal{K}\}$ have the following properties:
\begin{enumerate}
	\item Individual sparsity: The massive MIMO channels are usually correlated at the BTS side but not at the
	user side. This is due to the limited scattering at the BTS side, and relatively rich scattering at the users. Therefore, for a specific user $i$, the row vectors of $\bH^a_i$ usually have the same sparse support denoted as $\mathcal{S}_i$ with $0 < |\mathcal{S}_i| \ll M$, i.e., 
	\begin{equation}
	\label{eq3}
	\mathcal{S}_{i1} = \mathcal{S}_{i2} = \cdots = \mathcal{S}_{iN} \triangleq \mathcal{S}_i.
	\end{equation}
	\item Joint sparsity: Different users tend to share some common local scatterers at the BTS especially when the users are physically close to each other. We denote 
	\begin{equation}
	\label{eq4}
	\mathcal{C} = \bigcap^{K}_{i=1}\mathcal{S}_i.
	\end{equation}
\end{enumerate}

It is possible that users do not share common scatterers, i.e., $\mathcal{C} = \emptyset$. We show that the proposed algorithm remains valid using numerical results in Section \ref{sec:numerical}.

\begin{figure}[t]
	\centering
	\includegraphics[width=3.5in]{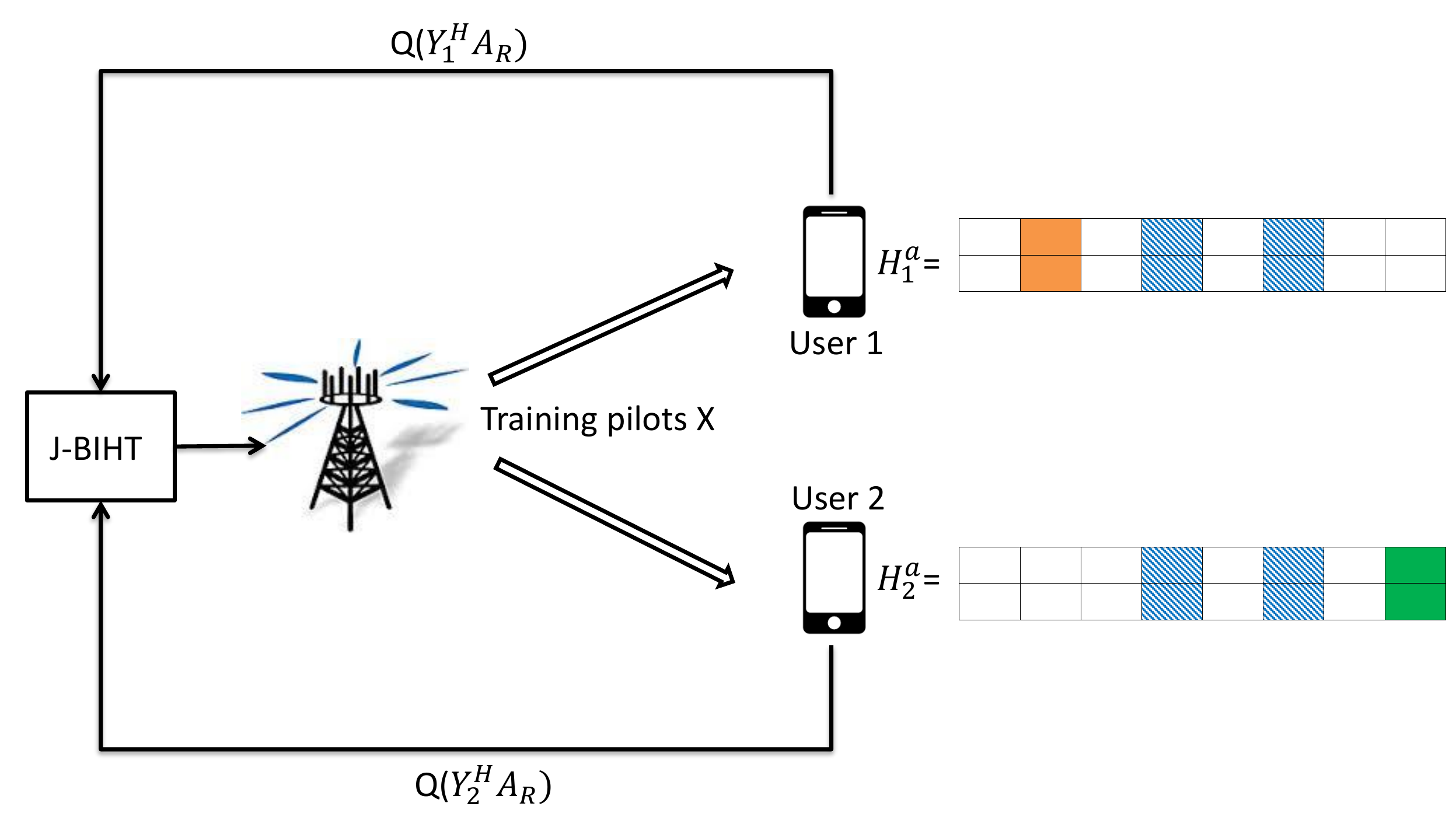}
	\caption{Schematic of the proposed CSIT estimation and one-bit feedback mechanism. The nonzero entries of the channel matrices are marked in color.}
	\label{Visio_system}
\end{figure}

Fig.~\ref{Visio_system} is an example of a two-user system, where the nonzero entries of the channel matrices are marked in color. The common support $ \mathcal{C}$ denotes the column indices corresponding to the blue shadowed entries. The support of user 1's channel $\mathcal{S}_1$ contains the column indices corresponding to the orange and blue shadowed entries of $\bH_1^a$. The support of user 2's channel $\mathcal{S}_2$ contains the column indices corresponding to the green and blue shadowed entries of $\bH_2^a$. 
 
Although the BTS has no prior knowledge about $\{\mathcal{S}_1,\mathcal{S}_2,\cdots,\mathcal{S}_K\}$ and $\mathcal{C}$, as in \cite{rao2014distributed}, we will assume that the BTS can measure bounds on the support defined as $\{s_i:i\in\mathcal{K}\}$ and $c$ such that $|\mathcal{C}|\geq c, |\mathcal{S}_i|\leq s_i,i\in\mathcal{K}$. Since these bounds depend on the large scale properties of the scattering environment which change slowly over time, they can be measured by the BTS over a slow timescale (e.g., seconds).

\section{CSIT Estimation: One-Bit Compressed Sensing}
\label{sec:channel}

\subsection{One-Bit Feedback Mechanism}

The CSIT estimation and one-bit feedback scheme is illustrated in Fig.~\ref{Visio_system}. Instead of estimating each $\bH_i$ at the $i$-th user based on the output symbols $\bY_i$, each user feeds back the one-bit quantized post-processed signal to the BTS. CSIT reconstruction is performed at the BTS using a joint recovery algorithm based on the feedback symbols. The procedure is described as Algorithm \ref{alg1}. 

\begin{algorithm}
	\caption{CSIT Estimation and One-Bit Feedback}
	\label{alg1}
	\begin{algorithmic}
		\STATE \textbf{Step 1. Pilot Training:} The BTS broadcasts training pilots $\bX\in \mathbb{C}^{M\times T}$ to all users;\\
		\STATE \textbf{Step 2. One-Bit Feedback:} \\
		The $i$-th user feeds back the one-bit quantized post-processed signal $Q(\bY_i^H\bA_R)$ to the BTS, where $Q(\bZ) = \text{sign}(\text{Re}(\bZ)) + j~\text{sign}(\text{Im}(\bZ))$.
		\STATE \textbf{Step 3. Joint CSIT Recovery at the BTS:} BTS recovers channel matrices $\{\bH_1^a,\bH_2^a,\cdots,\bH_K^a\}$ given $\{Q(\bY_1^H\bA_R),Q(\bY_2^H\bA_R),\cdots,Q(\bY_K^H\bA_R)\}$.
	\end{algorithmic}
\end{algorithm}

\subsection{Joint CSIT Recovery Algorithm}
There have been proposed compressed sensing algorithms for recovering jointly sparse signals \cite{cotter2005sparse,sarvotham2005distributed,fan2014enhanced,jin2015decentralized}. Our problem, however, is based on one-bit quantized measurements. In addition, the problem is more challenging than the conventional one-bit compressed sensing problem \cite{Boufounos2008bit} due to joint sparsity constraints (\ref{eq3}) and (\ref{eq4}). To make progress, we propose a joint binary iterative hard thresholding algorithm which takes advantage of the hidden joint sparsity structure of channel matrices. 

First, we rewrite (\ref{eq1}) into the standard one-bit compressed sensing model by defining the following transformations:
\begin{equation}
\hat{\bY}_i = Q(\bY_i^H\bA_R) \in \mathbb{C}^{T\times N}, \label{eq5}
\end{equation}
\begin{equation}
\hat{\bX} = \bX^H\bA_T \in \mathbb{C}^{T\times M}, \label{eq6}
\end{equation}
\begin{equation}
\hat{\bH}_i^a = (\bH_i^a)^H \in \mathbb{C}^{M\times N}, \label{eq7}
\end{equation}
\begin{equation}
\hat{\bN}_i = \bN_i^H\bA_R \in \mathbb{C}^{T\times N}. \label{eq8}
\end{equation} 
From (\ref{eq1}), we have
\begin{equation}
\hat{\bY}_i = Q(\hat{\bX}\hat{\bH}_i^a+\hat{\bN}_i), \forall i\in\mathcal{K}. \label{eq9}
\end{equation}
Equation (\ref{eq9}) is a standard one-bit compressed sensing measurement model, where $\hat{\bX}$ is the measurement matrix and $\hat{\bH}_i$ is the sparse matrix to be recovered. The amplitude of the signal has been lost during the quantization process. We restrict our attention to sparse signals on the unit sphere, i.e., our objective is to recover $\left\{\frac{\hat{\bH}_1}{{||\hat{\bH}_1||}_F}, \frac{\hat{\bH}_2}{{||\hat{\bH}_2||}_F},\cdots, \frac{\hat{\bH}_K}{{||\hat{\bH}_K||}_F}\right\}$, which is enough for many practical beamforming designs at the BTS. 

There have been several algorithms proposed to solve the one-bit compressed sensing problem. In \cite{Plan2013One}, the recovery problem is formulated as a convex programming. In \cite{Jacques2011Robust}, a greedy algorithm called Binary Iterative Hard Thresholding Algorithm (BIHT) is proposed. The objective of BIHT is to return a solution that is $k$-sparse and consistent with the given measurements. At each iteration, BIHT computes and takes a step in the direction of the gradient to attain a new approximation. This approximation is then projected onto the ``$\ell_0$ ball'', i.e., selecting the $K$ largest in magnitude element. Once the algorithm has terminated (either consistency is achieved or a maximum number of iterations have been reached), the final estimated signal is normalized by being projected onto the unit sphere. It is also extended from real-valued signals to complex-valued signals in \cite{stockle20151}. However, those methods do not directly apply to the recovery of CSIT in this paper due to the joint sparsity constraints (\ref{eq3}) and (\ref{eq4}). Therefore, we propose a modified joint binary iterative hard thresholding (J-BIHT) algorithm stated as Algorithm \ref{alg2} that can exploit the joint sparsity structure.

\begin{algorithm}
	\caption{J-BIHT}
	\label{alg2}
	\begin{algorithmic}
		\STATE \textbf{Input:} $\{\hat{\bY}_i:i\in\mathcal{K}\}, \bX, \{s_i:i\in\mathcal{K}\}, c, \mu$.
		\STATE \textbf{Step 1. Pre-processing:} Compute $\{\hat{\bY}_i:i\in\mathcal{K}\}, \hat{\bX}$ as in (\ref{eq5}), (\ref{eq6}).
		\STATE \textbf{Step 2. Initialization:} $\tilde{\bH}_i^a = \hat{\bX}^H\hat{\bY}_i, k=0$.
		\STATE \textbf{Step 3. Update:} $\tilde{\bH}_i^a\leftarrow \tilde{\bH}_i^a - \mu\hat{\bX}^H(Q(\hat{\bX}\tilde{\bH}_i^a)-\hat{\bY}_i)$,  
		\begin{itemize}
			\item $\tilde{\mathcal{S}}_i \leftarrow \mathop{\argmax}\limits_{\substack{\mathcal{I}\subset\{1,2,\cdots,M\} \\ \text{card}(\mathcal{I})=s_i}} \sum\limits_{j\in\mathcal{I}}{||\tilde{\bh}_{ij}||}^2_2$, where $\tilde{\bh}_{ij}$ is the $j$-th row of $\tilde{\bH}_i^a$ and $\text{card}(\mathcal{I})$ returns the cardinality of set $\mathcal{I}$.\\
			\item $\tilde{\mathcal{C}} \leftarrow \mathop{\argmax}\limits_{\substack{\mathcal{I}\subset\{1,2,\cdots,M\} \\ \text{card}(\mathcal{I})=c}} \sum\limits_{j\in\mathcal{I}}\text{mode}(j,\{\tilde{\mathcal{S}}_1,\tilde{\mathcal{S}}_2,\cdots,\tilde{\mathcal{S}}_K\})$, where $\text{mode}(j,\mathcal{I})$ returns the number of times that $j$ appears in set $\mathcal{I}$.\\
			\item $\tilde{\mathcal{S}}_i \leftarrow \tilde{\mathcal{C}} \cup \mathop{\argmax}\limits_{\substack{\mathcal{I}\subset\{1,2,\cdots,M\} \\ \text{card}(\mathcal{I})=s_i-c \\ \mathcal{I} \cap \tilde{\mathcal{C}} = \emptyset}} \sum\limits_{j\in\mathcal{I}}{||\tilde{\bh}_{ij}||}^2_2$.   \\
			\item Hard threshold all but the entries in $\tilde{\mathcal{S}}_i$.
		\end{itemize}
		
		\STATE \textbf{Step 4.} If the estimated $\{\tilde{\bH}_i^a,i\in\mathcal{K}\}$ are consistent with $\{\hat{\bY}_i,i\in\mathcal{K}\}$ or iteration count is sufficient, compute $\tilde{\bH}_i = \bA_R(\tilde{\bH}_i^a)^H\bA_T^H$ and
		return $\left\{\frac{\tilde{\bH}_1}{{||\tilde{\bH}_1||}_F}, \frac{\tilde{\bH}_2}{{||\tilde{\bH}_2||}_F},\cdots, \frac{\tilde{\bH}_K}{{||\tilde{\bH}_K||}_F}\right\}$; otherwise, go back to Step 3.
	\end{algorithmic}
\end{algorithm}
Note that $\mu$ is the step size for gradient descent. In Algorithm \ref{alg2}, Step 3 aims to identify the common support first, then the partial support side information $\tilde{\mathcal{C}}$ is used to improve the CSIT estimation performance of each $\tilde{\bH}_i^a$.

\subsection{Design of Training Pilots} 
The design of training pilots is crucial to the performance of the proposed scheme. In \cite{gupta2010sample}, the measurement
matrix is assumed to be Gaussian and it is demonstrated that the support of an $n$-dimensional $s$-sparse signal can tractably be recovered from $O(s\log n)$ measurements.
In \cite{Jacques2011Robust}, a certain \textit{binary-stable embedding property} which is a one-bit analogue
to the restricted isometry property of standard compressed sensing is introduced. It is also shown that Gaussian
measurement ensembles satisfy this property with high probability (given enough measurements). In particular, $O(s \log n)$ Gaussian measurements are sufficient to have a relative error bounded by any fixed constant. These results are
robust to noise. Therefore, the training pilots $\bX \in \mathbb{C}^{M\times T}$ can be designed as $\bX = \bA_T\bZ$, where $\bZ\in\mathbb{C}^{M\times T}$ consists of independent identically distributed random variables, each is drawn from $\left\{-\sqrt{\frac{P}{M}},\sqrt{\frac{P}{M}}\right\}$.

\subsection{Overhead}
The main motivation for our scheme is to reduce the number of bits as opposed to the number of measurements. Although channel estimation using conventional compressed sensing helps to reduce the number of measurements required to recover the signal, in the presence of coarsely quantized measurements (feedback), the performance may not be adequate. Although the performance can be improved by adding measurements or using a more precise quantizer, the total information overhead (including training overhead and feedback overhead) will increase relative to the feedback overhead in the proposed scheme.

\section{Numerical Results}
\label{sec:numerical}

In this section, we illustrate the performance gain of the proposed J-BIHT algorithm by comparing it with the following baseline algorithms:
\begin{enumerate}
	\item BIHT: Each $\bH_i$ is recovered individually from the feedback $Q(\bY_i^H\bA_R)$ using the BIHT algorithm \cite{Jacques2011Robust}.
	\item J-BIHT given knowledge of the common support: The channel matrices $\{\bH_i:i\in\mathcal{K}\}$ are jointly recovered using the proposed J-BIHT algorithm given the support information $\{\mathcal{S}_i:i\in\mathcal{K}\}$ and $\mathcal{C}$, which acts as a performance upper bound on the proposed J-BIHT algorithm.
	\item Genie-aided Least Squares (LS): We assume the BTS has the support information $\{\mathcal{S}_i:i\in\mathcal{K}\}$ and $\mathcal{C}$. Furthermore, perfect feedback $\{\bY_i:i\in\mathcal{K}\}$ from each user is also assumed such that the channel matrices $\{\bH_i:i\in\mathcal{K}\}$ are jointly recovered directly using LS.
\end{enumerate}

Consider a multi-user massive MIMO system with $M = 128$ antennas at the BTS and $N = 2$ antennas at each user. There are $K = 10$ users. The individual sparsity levels $|\mathcal{S}_i|$ are independently generated from a uniform distribution over $\{s-2,s-1,s\}$. The joint sparsity level is generated from a uniform distribution over $\{c,c+1\}$. The SNR at the BTS is set to be 15 dB. The step size $\mu$ in Algorithm \ref{alg2} is set to be $\mu = 0.01$.

The metric we use for comparison is the average output SNR degradation. Given $\bH_i$, the optimal precoder at the BTS denoted as ${\bw}_i\in \mathbb{C}^{M\times 1}$ is the normalized eigenvector corresponding to the largest eigenvalue of $\bH_i^H \bH_i$, i.e., ${\bw}_i$ is the maximizer of $||\bw_i^H {\bH}_i^H {\bH}_i \bw_i||_2^2$ with ${||{\bw}_i||}_2=1$. The precoder $\tilde{\bw}_i$ based on the estimated CSIT is computed as the maximizer of $||\tilde{\bw}_i^H \tilde{\bH}_i^H \tilde{\bH}_i \tilde{\bw}_i||_2^2$ with ${||\tilde{\bw}_i||}_2=1$.
For each channel realization, the SNR loss in dB is calculated as 
\begin{align}
SNR_{loss} = 10 \log_{10} \frac{||\bw_i^H \bH_i^H \bH_i \bw_i||_2^2}{||\tilde{\bw}_i^H \bH_i^H \bH_i \tilde{\bw}_i||_2^2}.
\end{align}
In the simulation results, the average output SNR degradation is obtained by averaging over 100 random channel realizations.

\begin{figure}[t]
	\centering
	\includegraphics[width=3.5in]{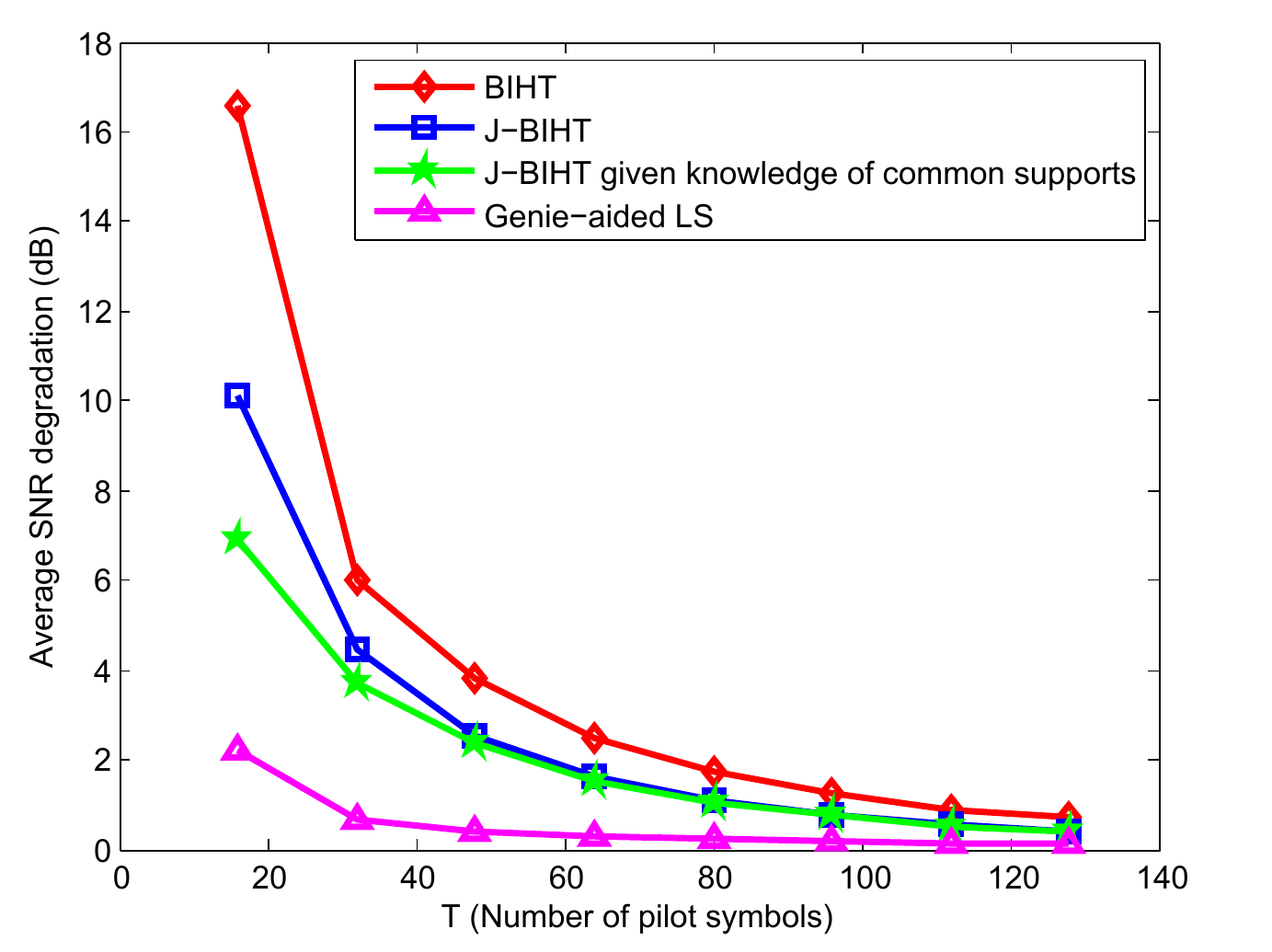}
	\caption{Average output SNR degradation versus the number of pilot symbols $T$.}
	\label{T}
\end{figure}

In Fig. \ref{T}, we compare the average output SNR degradation versus the number of training as well as feedback overhead $T$, under the individual sparsity parameter $s = 10$ and joint sparsity parameter $c = 6$. Due to the utilization of the joint sparsity structure, it is clear that the proposed J-BIHT algorithm yields significant gains compared to using BIHT algorithm individually. It is also shown that, as the number of training pilots increases, J-BIHT achieves exactly the same gain as if the knowledge of common supports is given. This means J-BIHT can exactly identify all the common supports given enough one-bit measurements. Moreover, it is also observed that the gain achieved by J-BIHT is close to the optimal genie-aided LS and it approaches the optimum as the number of training pilots increases. When the feedback overhead is as small as about 64 bits, the output SNR degradation by using J-BIHT is already less than 2 dB. This is substantial compared to using quantized feedback which has large overhead. 

\begin{figure}[t]
	\centering
	\includegraphics[width=3.5in]{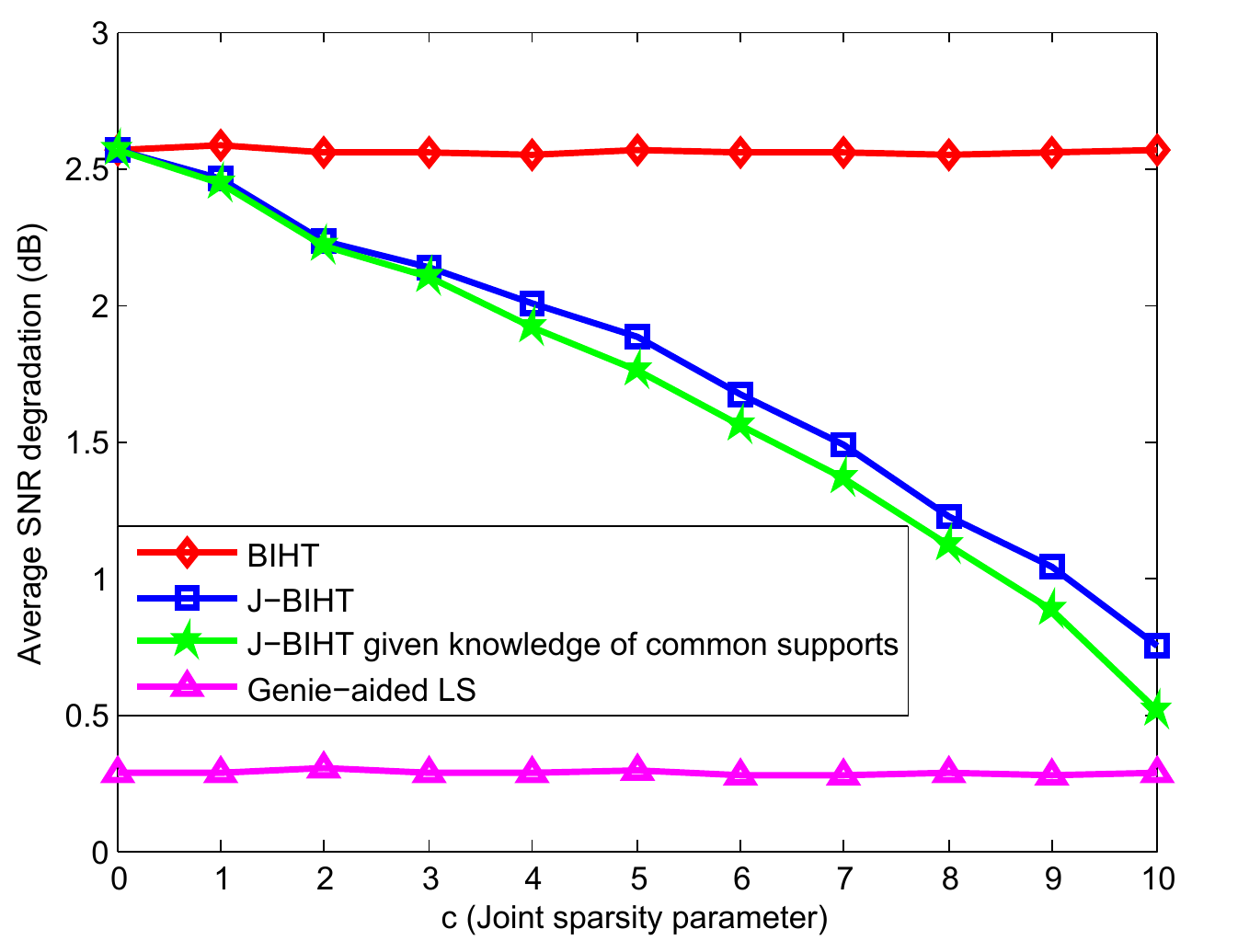}
	\caption{Average output SNR degradation versus the joint sparsity parameter $c$.}
	\label{Sc}
\end{figure}

Fig.~\ref{Sc} compares the average output SNR degradation versus the joint sparsity parameter $c$ under the number of training pilots $T=64$ and individual sparsity parameter $s = 10$. Fig.~\ref{Sc} shows that the performance gets better as the number of common supports increases. This is because the proposed J-BIHT exploits the joint sparsity structure of channel matrices. Therefore, all individual supports are more likely to be recovered given more knowledge of the common supports. Moreover, when $c = 0$, i.e., users do not share common scatterers, and J-BIHT loses about 2.5 dB in output SNR. It is possible to improve the performance by clustering users who are close to each other and so are more likely to share common scatterers, and running J-BIHT for each cluster.

\begin{figure}[t]
	\centering
	\includegraphics[width=3.5in]{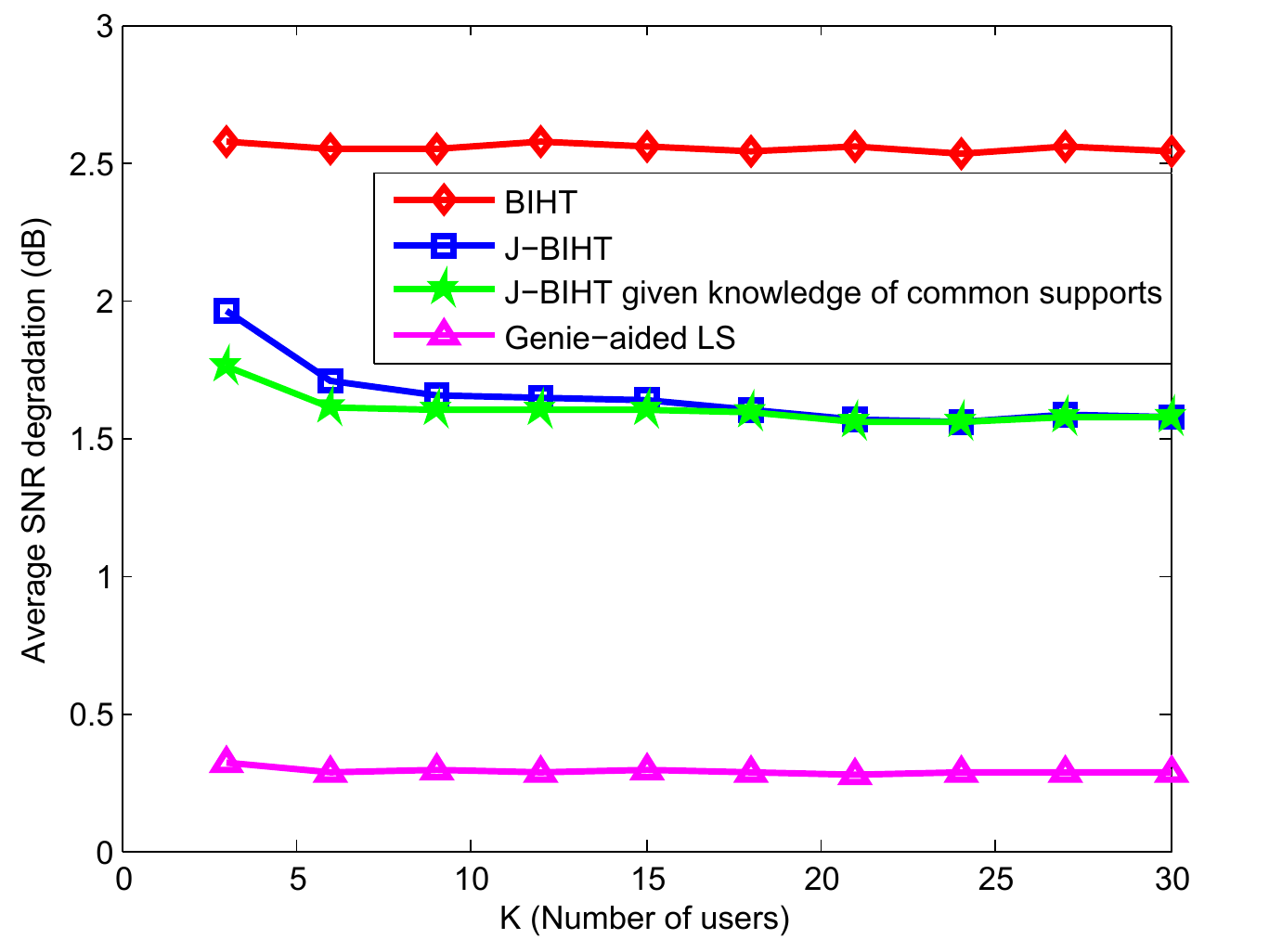}
	\caption{Average output SNR degradation versus the number of users $K$.}
	\label{K}
\end{figure}

Fig. \ref{K} compares the average output SNR degradation versus the number of users $K$ under the number of training pilots $T=64$, individual sparsity parameter $s = 10$ and joint sparsity parameter $c = 6$. From Fig. \ref{K}, we can observe that as the number of users increases, the proposed J-BIHT yields better performance and the common supports are identified more accurately. This is because J-BIHT is aware of the joint sparsity structure of all channel matrices. Therefore, more users gives better estimation of the common supports, which leads to better estimation of all channel matrices. However, the gain achieved using BIHT individually is not improved due to the lack of joint sparsity information.

\begin{figure}[t]
	\centering
	\includegraphics[width=3.5in]{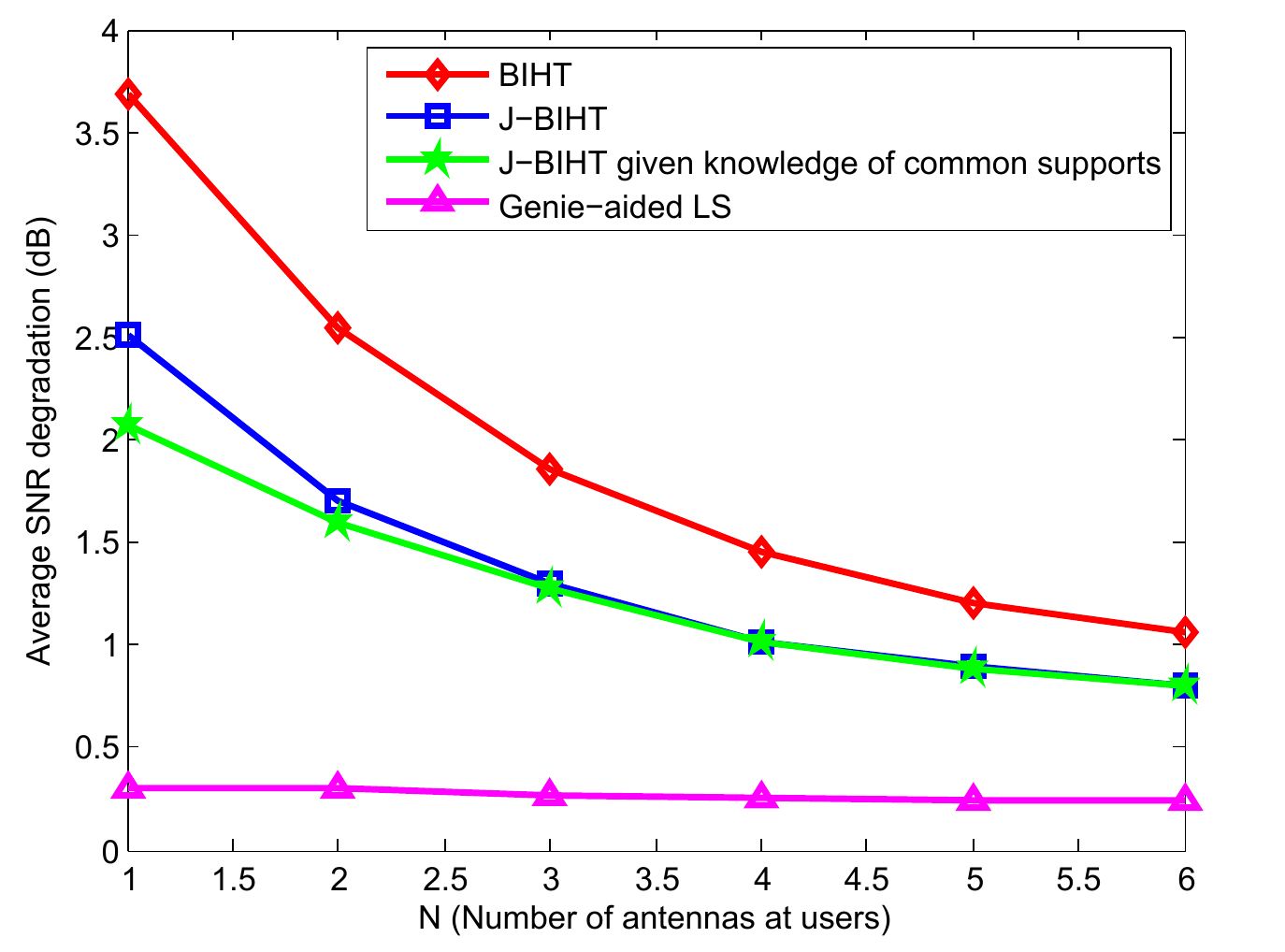}
	\caption{Average output SNR degradation versus $N$ (the number of antennas at users).}
	\label{N}
\end{figure}

Fig. \ref{N} compares the average output SNR degradation versus $N$ (the number of antennas at users) with the number of training pilots $T=64$, individual sparsity parameter $s = 10$ and joint sparsity parameter $c = 6$. It is observed that as the number of each user antennas increases, the proposed J-BIHT yields better performance and the common supports are identified more accurately. This is because the individual sparsity among all $N$ row vectors of each channel matrix is exploited by J-BIHT. Therefore, larger $N$ gives better estimation quality. Moreover, as $N$ increases, the performance of J-BIHT becomes much closer to that of optimal genie-aided LS, which shows the potential of J-BIHT for optimal beamforming with only one bit for each feedback symbol.

\section{Conclusion}
\label{sec:conclude}
This paper has studied downlink channel estimation in multi-user massive MIMO systems. A practical one-bit compressed sensing based scheme has been proposed to greatly reduce the training and feedback overhead. A joint sparsity recovery algorithm has been proposed to accurately estimate the channel matrices for the beamforming design at the BTS. Numerical results show that, by taking into account the joint sparsity structure of channel matrices, the proposed scheme can achieve close-to-optimal performance (with less than 0.5 dB loss in terms of output SNR) with a small amount of training/feedback overhead. A possibility for future work is to characterize the performance of the proposed scheme analytically. Other ongoing work is to design an algorithm which adapts the number of training pilots with the estimated partial knowledge.





\bibliographystyle{IEEEtran}
\bibliography{IEEEabrv,Zhiyi_bib,all_bib}

\end{document}